\pdfoutput=1
\documentclass[journal]{IEEEtran}

\RequirePackage[utf8]{inputenc}
\RequirePackage[english]{babel}
\RequirePackage{amsmath,amsfonts}
\RequirePackage{algorithmic}
\RequirePackage[pscoord]{eso-pic}
\RequirePackage{textcomp}
\PassOptionsToPackage{detect-all, per-mode=symbol, range-units=single, range-phrase=~to~}{siunitx}
\RequirePackage{siunitx}
\RequirePackage{nicefrac}
\RequirePackage[hidelinks]{hyperref}
\RequirePackage{orcidlink}
\RequirePackage[noabbrev,capitalise]{cleveref}
\RequirePackage[caption=false]{subfig}
\RequirePackage{url}
\RequirePackage{lipsum}
\RequirePackage{placeins}
\RequirePackage{xfrac}
\RequirePackage{float}

\RequirePackage{amssymb}
\RequirePackage{pifont}

\RequirePackage{threeparttable}
\RequirePackage{booktabs}
\RequirePackage{colortbl}
\RequirePackage{tabularx}
\RequirePackage{makecell}

\RequirePackage{graphicx}
\RequirePackage{glossaries}
\RequirePackage{xspace}
\RequirePackage{lipsum}
\RequirePackage{microtype}
\RequirePackage{multirow}

\newcommand{\glsf}[1]{\glsreset{#1}\gls{#1}}
\newcommand{\glsplf}[1]{\glsreset{#1}\glspl{#1}}

 \usepackage{colortbl}

\renewcommand{\subsection}[1]{\paragraph*{\textbf{#1}}}
\renewcommand{\subsubsection}[1]{\paragraph*{\textbf{#1}}}

\microtypesetup{expansion=true, protrusion=true}
\begin{document}

\newacronym{sota}{SOTA}{state-of-the-art}
\newacronym{fpga}{FPGA}{field programmable gate array}
\newacronym{asic}{ASIC}{application-specific integrated circuit}
\newacronym{fub}{FUB}{functional unit block}
\newacronym{vv}{V\&V}{validation and verification}
\newacronym{gpp}{GPP}{general purpose processor}
\newacronym{tid}{tID}{transaction ID}
\newacronym{cots}{COTS}{commercial-off-the-shelf}

\newacronym{hpc}{HPC}{high performance computing}
\newacronym{ml}{ML}{machine learning}
\newacronym{isa}{ISA}{instruction set architecture}
\newacronym{fp}{FP}{floating-point}
\newacronym{dl}{DL}{deep learning}
\newacronym{la}{LA}{linear algebra}
\newacronym{ip}{IP}{intellectual property}
\newacronym[firstplural=systems-on-chip (SoCs)]{soc}{SoC}{system-on-chip}
\newacronym{mpsoc}{MPSoC}{multi-processor system-on-chip}
\newacronym[firstplural=networks-on-chip (NoCs)]{noc}{NoC}{network-on-chip}
\newacronym{hw}{HW}{hardware}
\newacronym{sw}{SW}{software}
\newacronym{swapc}{SWaP-C}{space, weight, power, and cost}
\newacronym{mcp}{MCP}{multi-core processor}
\newacronym{rr}{RR}{round-robin}
\newacronym{tdma}{TDMA}{time division multiple access}
\newacronym{aces}{ACES}{Autonomous driving, Connectivity, Electrification, and Shared mobility} 
\newacronym{sdv}{SDV}{software-defined vehicles}

\newacronym{mac}{MAC}{multiply-accumulate}
\newacronym{fem}{FEM}{finite element analysis}
\newacronym{simd}{SIMD}{single-instruction, multiple-data}
\newacronym{rtl}{RTL}{register transfer level}
\newacronym{dlt}{DLT}{data layout transform}

\newacronym{rv}{RV}{RISC-V}
\newacronym{rvu}{RVVU}{RISC-V Vector Unit}
\newacronym{cc}{CC}{core complex}
\newacronym{secd}{SECD}{secure domain}
\newacronym{safed}{SAFED}{safe domain}
\newacronym{hostd}{HOSTD}{host domain}
\newacronym{amr}{AMR}{adapative modular redundancy}
\newacronym{hfr}{HFR}{hardware fast recovery}
\newacronym{vecd}{VECD}{vector cluster domain}
\newacronym{rot}{RoT}{root-of-trust}
\newacronym{dc}{D\$}{data cache}
\newacronym{dpllc}{DPLLC}{dynamic partitionable last-level cache}
\newacronym{bw}{BW}{bandwidth}
\newacronym{dcspm}{DCSPM}{dynamically configurable L2 scratchpad memory}
\newacronym{mct}{MCT}{mixed-criticality task}
\newacronym{mc}{MC}{mixed critical}
\newacronym{tct}{TCT}{time-critical task}
\newacronym{tsu}{TSU}{traffic shaper unit}
\newacronym{gbs}{GBS}{granular burst splitter}
\newacronym{wb}{WB}{write buffer}
\newacronym{tru}{TRU}{traffic regulation unit}
\newacronym{nct}{NCT}{non-critical task}
\newacronym{vg}{VG}{virtual guest}
\newacronym{gemm}{GEMM}{general matrix-multiply}
\newacronym{matmul}{MatMul}{matrix multiplication}
\newacronym{indip}{INDIP}{independent mode}
\newacronym{ecc}{ECC}{error correction code}
\newacronym{pcb}{PCB}{printed circuit board}
\newacronym{dlm}{DLM}{dual-lockstep mode}
\newacronym{tlm}{TLM}{triple-lockstep mode}
\newacronym{ffts}{FFTs}{fast Fourier transforms}
\newacronym{vrf}{VRF}{vector register file}
\newacronym{sdotp}{sdotp}{sum-of-dot-product}
\newacronym{ai}{AI}{artificial intelligence}
\newacronym{vau}{VAU}{Vector Arithmetic Unit}
\newacronym{dsp}{DSP}{Digital Signal Processing}
\newacronym{soa}{SoA}{state-of-the-art}
\newacronym{gp}{GP}{general purpose}

\newacronym{fifo}{FIFO}{first in, first out}
\newacronym{fu}{FU}{functional unit}
\newacronym{alu}{ALU}{arithmetic logic unit}
\newacronym{fpu}{FPU}{floating-point unit}
\newacronym{ssr}{SSR}{stream semantic register}
\newacronym{issr}{ISSR}{indirection stream semantic register}
\newacronym{tcdm}{TCDM}{tightly-coupled data memory}
\newacronym{dma}{DMA}{direct memory access}
\newacronym{sm}{SM}{streaming multiprocessor}
\newacronym{vlsu}{VLSU}{vector load-store unit}
\newacronym{dsa}{DSA}{domain-specific accelerator}
\newacronym{ha}{HA}{hardware accelerator}
\newacronym{fsm}{FSM}{finite state machine}
\newacronym{llc}{LLC}{last-level cache}
\newacronym{d2d}{D2D}{die-to-die}
\newacronym{dram}{DRAM}{dynamic random access memory}
\newacronym{spm}{SPM}{scratchpad memory}
\newacronym{rf}{RF}{register file}
\newacronym{mmu}{MMU}{memory management unit}
\newacronym{pmp}{MMU}{physical memory protection unit}
\newacronym{l2}{L2}{level-two}
\newacronym{clic}{CLIC}{core-local interrupt controller}
\newacronym{vclic}{vCLIC}{virtualized CLIC}
\newacronym{eth}{ETH}{Ethernet}

\newacronym{os}{OS}{operating system}

\newacronym{spvv}{SpVV}{sparse vector-vector multiplication}
\newacronym{spmv}{SpMV}{sparse vector-matrix multiplication}
\newacronym{spmm}{SpMM}{sparse matrix-matrix multiplication}
\newacronym{csrmv}{CsrMV}{CSR matrix-vector multiplication}
\newacronym{csrmm}{CsrMM}{CSR matrix-matrix multiplication}

\newacronym{csf}{CSF}{compressed sparse fiber}
\newacronym{csr}{CSR}{compressed sparse rows}
\newacronym{csc}{CSC}{compressed sparse columns}
\newacronym{bcsr}{BCSR}{blocked compressed sparse rows}

\newacronym{axi4}{AXI4}{Advanced eXtensible Interface 4}
\newacronym{amba}{AMBA}{Advanced Microcontroller Bus Architecture}
\newacronym{sram}{SRAM}{static random-access memory}

\newacronym{wcet}{WCET}{worst-case execution time}
\newacronym{wcdt}{WCDT}{worst-case detection time}
\newacronym{rtunit}{REALM unit}{real-time regulation and traffic monitoring unit}
\newacronym{mtunit}{M\&R unit}{monitoring and regulation unit}
\newacronym{cps}{CPS}{cyber-physical system}
\newacronym{crtes}{CRTES}{critical real-time embedded system}
\newacronym{heicps}{He-iCPS}{heterogeneous integrated cyber-physical system}
\newacronym{ecu}{ECU}{electronic control unit}
\newacronym{mcs}{MCS}{mixed criticality system}
\newacronym{ima}{IMA}{integrated modular avionics}
\newacronym{adas}{ADAS}{Advanced Driver Assistance System} 
\newacronym{axirealm}{AXI-REALM}{AXI real-time regulation and traffic monitoring}
\newacronym{mpam}{MPAM}{memory system resource partitioning and monitoring}
\newacronym{dos}{DoS}{denial of service}
\newacronym{hwrot}{HWRoT}{hardware root of trust}
\newacronym{pcs}{PCS}{power controller subsystem}
\newacronym{sil}{SIL}{safety integrity level}
\newacronym{tmu}{TMU}{transaction monitor unit}

\newacronym{ps}{PS}{per-system}
\newacronym{pu}{PU}{per-unit}
\newacronym{pur}{PUR}{per unit and region}

\newcommand{\gf}{{GlobalFoundries}}
\newcommand{\gfs}{{GlobalFoundries'}}
\newcommand{\gftech}{{GF12LP+}}
\newcommand{\dc}{{Synopsys} {Design} {Compiler} {NXT} 2023.12}
\newcommand{\pt}{{Synopsys} {PrimeTime} 2022.03}
\newcommand{\riscv}{\mbox{RISC-V}}

\title{A Reliable, Time-Predictable Heterogeneous SoC for AI-Enhanced Mixed-Criticality Edge Applications}

\author{
    Angelo Garofalo~\orcidlink{0000-0002-7495-6895}\IEEEauthorrefmark{10}, 
    Alessandro Ottaviano~\orcidlink{0009-0000-9924-3536}\IEEEauthorrefmark{10}, 
    Matteo Perotti~\orcidlink{0000-0003-2413-8592}, 
    Thomas Benz~\orcidlink{0000-0002-0326-9676}, 
    Yvan Tortorella~\orcidlink{0000-0001-8248-5731}, 
    Robert Balas~\orcidlink{0000-0002-7231-9315}, 
    Michael Rogenmoser~\orcidlink{0000-0003-4622-4862}, 
    Chi Zhang~\orcidlink{0009-0003-3243-0558}, 
    Luca Bertaccini~\orcidlink{0000-0002-3011-6368}, 
    Nils Wistoff~\orcidlink{0000-0002-8683-8060}, 
    Maicol Ciani~\orcidlink{0009-0003-7861-9129}, 
    Cyril Koenig~\orcidlink{0009-0007-9771-1691}, 
    Mattia Sinigaglia~\orcidlink{0000-0002-3350-8789}, 
    Luca Valente~\orcidlink{0000-0002-7458-477X}, 
    Paul Scheffler~\orcidlink{0000-0003-4230-1381}, 
    Manuel Eggimann~\orcidlink{0000-0001-8395-7585}, 
    Matheus Cavalcante~\orcidlink{0000-0001-9199-1708}, 
    Francesco Restuccia~\orcidlink{0000-0001-6955-1888}, 
    Alessandro Biondi~\orcidlink{0000-0002-6625-9336}, 
    Francesco Conti~\orcidlink{0000-0002-7924-933X}, 
    Frank K. Gurkaynak~\orcidlink{0000-0002-8476-554X}, 
    Davide Rossi~\orcidlink{0000-0002-0651-5393}, 
    Luca Benini~\orcidlink{0000-0001-8068-3806}
    \vspace{-0.5cm}
    \IEEEcompsocitemizethanks{%
    \IEEEauthorrefmark{10} Both authors contributed equally to this research.\protect\\
    \IEEEcompsocthanksitem A.~Garofalo, A.~Ottaviano, M.~Perotti, T.~Benz, R.~Balas, M.~Rogenmoser, C.~Zhang, L.~Bertaccini, N.~Wistoff,  C.~Koenig, P.~Scheffler, M.~Eggimann, M.~Cavalcante, F.~K.~Gurkaynak,  and L.~Benini are with the D-ITET department at ETH Zurich, Switzerland.\protect\\
    E-mail: \{agarofalo,aottaviano\}@ethz.ch\protect\\
    Y.~Tortorella, M.~Ciani, M.~Sinigaglia, L.~Valente, F.~Conti, D.~Rossi are with the Department of Electrical, Electronic, and Information Engineering (DEI), University of Bologna, Italy.\protect\\ 
    F.~Restuccia is with the Department of Computer Science and Engineering, University of California in San Diego (UCSD), California, USA.\protect\\ 
    A.~Biondi is with the Department of Excellence in Robotics \& AI, Scuola Superiore Sant'Anna, Pisa, Italy.\protect\\
    }
}
\markboth{Journal of \LaTeX\ Class Files,~Vol.~14, No.~8, August~2021}%
{Shell \MakeLowercase{\textit{et al.}}: A Sample Article Using IEEEtran.cls for IEEE Journals}

\maketitle

\begin{abstract}
Next-generation mixed-criticality Systems-on-chip (SoCs) for robotics, automotive, and space must execute mixed-criticality AI-enhanced sensor processing and control workloads, ensuring reliable and time-predictable execution of critical tasks sharing resources with non-critical tasks, while also fitting within a sub-2W power envelope. To tackle these multi-dimensional challenges, in this brief, we present a 16nm, reliable, time-predictable heterogeneous SoC with multiple programmable accelerators. Within a 1.2W power envelope, the SoC integrates software-configurable hardware IPs to ensure predictable access to shared resources, such as the on-chip interconnect and memory system, leading to tight upper bounds on execution times of critical applications. To accelerate mixed-precision mission-critical AI, the SoC integrates a reliable multi-core accelerator achieving 304.9 GOPS peak performance at 1.6 TOPS/W energy efficiency. Non-critical, compute-intensive, floating-point workloads are accelerated by a dual-core vector cluster, achieving 121.8 GFLOPS at 1.1 TFLOPS/W and 106.8 GFLOPS/mm$^2$.
\end{abstract}


\section{Introduction}

To meet the growing computational demands of \gls{ai}-enhanced applications in safety-critical domains like automotive, space, and robotics, near-sensors zonal/on-board controllers~\cite{burkacky2023getting} must efficiently handle compute-intensive tasks while ensuring reliable, time-predictable execution of \glspl{tct}~\cite{rehmRoadPredictableAutomotive2021}. To optimize latency, performance, efficiency, and area, Systems-on-chip (SoCs) must be designed as heterogeneous \glspl{mcs}~\cite{jiangReThinkingMixedCriticalityArchitecture2020}, combining \gls{gp} processors and \glspl{dsa}, within a sub-2W power budget typical of high-end microcontrollers~\cite{ojoReviewLowEndMiddleEnd2018}. 

These multi-faceted requirements pose significant design challenges. For instance, achieving time predictability without major performance overhead is hindered by resource conflicts among heterogeneous compute units sharing interconnects and memory endpoints~\cite{majumder2020partaa}, making it difficult to ensure a bounded \gls{wcet}. \textit{Spatial} and \textit{temporal} partitioning of hardware resources~\cite{kloda2019deterministic} are key techniques to mitigate these issues. While recent works partially address this problem in hardware, mainly focusing on interconnects~\cite{jiang2022axi, benz2025axi}, no comprehensive hardware implementation providing observability and controllability of shared resources for predictability has been demonstrated in silicon. As a result, freedom-from-interference or bounded \gls{wcet} on \gls{soa} SoCs~\cite{nxp_industrial_control, valenteHeterogeneousRISCVBased2024, grossierASILDAutomotivegradeMicrocontroller2023} is mainly achieved through software mechanisms, leading to significant performance overhead~\cite{6531078}.

\begin{figure}[!t]
    \centering
    \includegraphics[width=0.98\columnwidth]{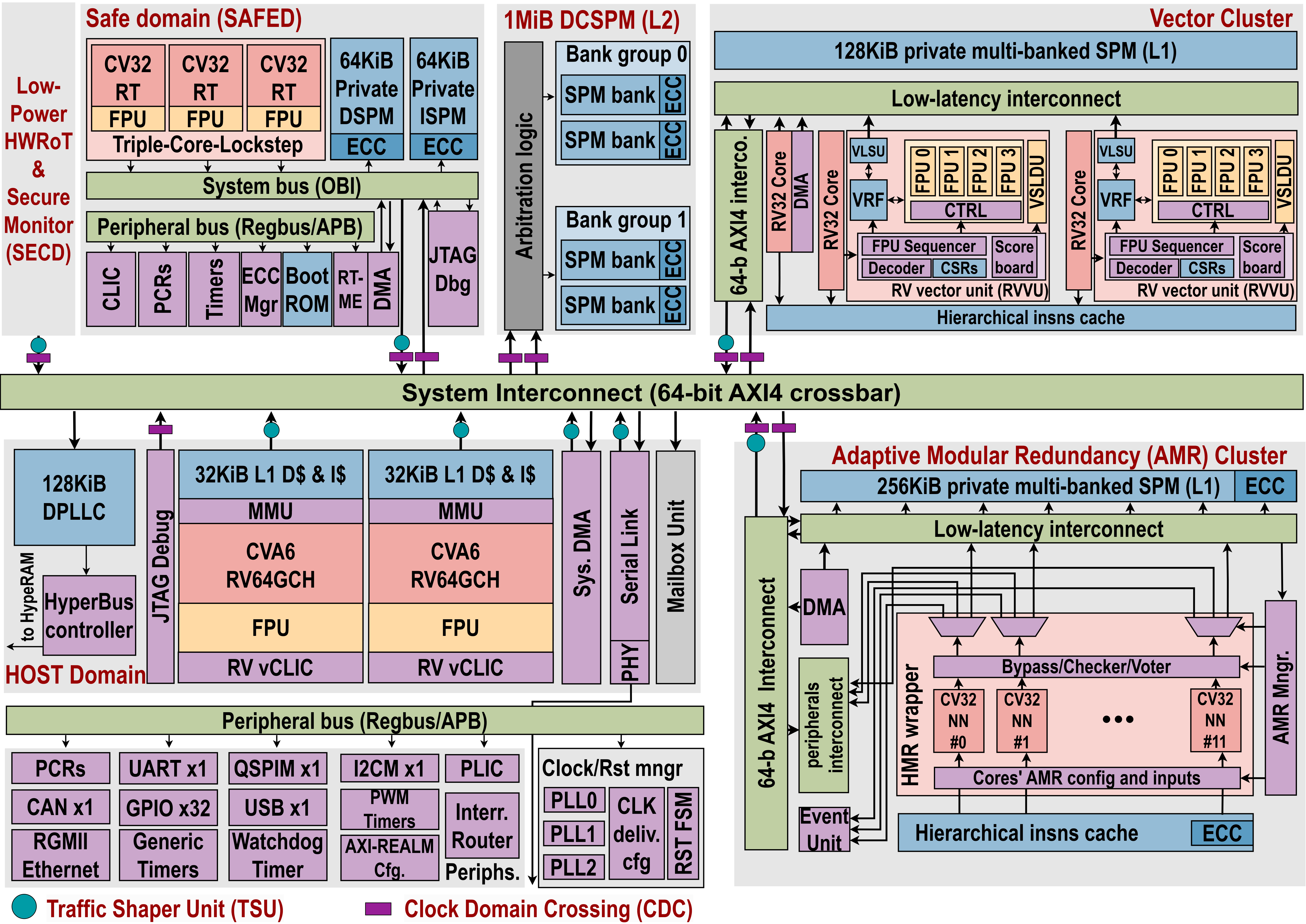}
    \caption{SoC Architecture.}
    \label{fig:soc-archi}
\end{figure}

\begin{figure*}[!t]
    \centering
    \includegraphics[width=0.99\linewidth]{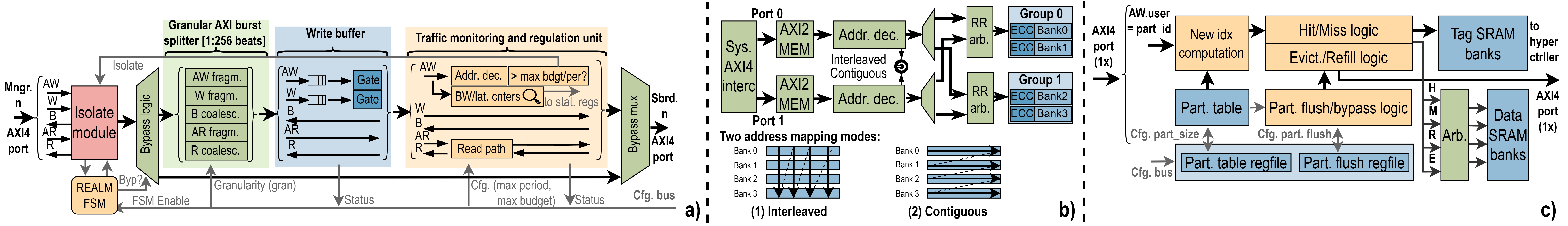}
    \caption{Architectures of the hardware IPs for predictability: a) \gls{tsu}; b) \gls{dcspm}; c) \gls{dpllc}.}
    \label{fig:hw-ips-for-predictability}
\end{figure*}

Another challenge is ensuring that \glspl{dsa} meet both performance and dependability requirements while remaining compact and low-power.
On the one hand, traditional high-precision (\gls{fp}) tasks for \gls{dsp} (e.g., radar signal processing) and advanced predictive control require acceleration to meet stringent control-loop bandwidth constraints~\cite{islayem2024hardware, jhung2023hardware}. 
On the other hand, a new class of \gls{ai}-intensive low arithmetic precision tasks (e.g. deep neural network inference for object detection, collision avoidance, condition monitoring) require not only tight real-time constraints but also fault tolerance and resiliency~\cite{islayem2024hardware}.

Despite domain-specific acceleration (\gls{dsp}-control, \gls{ai}-perception) being crucial for performance and energy efficiency, software programmability remains a fundamental pillar for keeping pace with rapidly evolving algorithms. The extensible nature of the \gls{rv} \gls{isa} with custom instructions offers a viable approach to specialize instruction processors, striking a balance between efficiency, performance, and programmability~\cite{valenteHeterogeneousRISCVBased2024, UMBRELLA_TCASI_1}. However, while recent research on programmable edge \gls{ai} processors has mainly focused on performance and efficiency metrics~\cite{valenteHeterogeneousRISCVBased2024, xu2023ultra, sun202340nm, juSystolicNeuralCPU2023}, there is a need for silicon-proven, dependable hardware architectures achieving \gls{soa} performance and energy efficiency.



To the best of our knowledge, no SoC addresses these challenges holistically. To close this gap, we present a heterogeneous \gls{rv} SoC implemented in Intel 16nm FinFet technology, featuring three main contributions: 
\textbf{1)} end-to-end time-predictable execution of \glspl{mct} enabled by a set of software-programmable hardware IPs for configurable partitioning of shared resources at zero performance overhead, namely: a configurable \glsf{tsu} for interconnect, a \glsf{dpllc}, and the L2 \glsf{dcspm}; 
%
%
\textbf{2)} a 1.07 TFLOPS/W, 107 GFLOPS/mm$^2$-at-FP8 dual-core vector floating-point mixed-precision acceleration cluster for \gls{fp} workloads; 
\textbf{3)} a 1.61 TOPS/W, 260.7 GOPS/mm$^2$ at 2b 12-core \gls{rv} integer accelerator cluster with runtime \gls{amr} to trade-off performance with reliability in integer mixed-precision mission-critical \gls{ai} tasks. To enable extensions, benchmarking and comparative analysis, we release the synthesizable hardware description of the SoC design open-source under a liberal license~\footnote{\texttt{\url{https://github.com/pulp-platform/carfield}}}. 

\section{SoC Architecture}
Fig.~\ref{fig:soc-archi} shows the SoC architecture, operating across three clock domains, each driven by a dedicated PLL. A \gls{secd} acts as the SoC’s \gls{hwrot}, handling the secure boot and crypto services. 
For hard real-time and safety-critical tasks, the \gls{safed} features a lockstepped triple-RV32-core for reliability, \gls{ecc}-protected private instruction and data \gls{spm} for deterministic memory access, and an enhanced \gls{rv} \gls{clic} with 6-cycle interrupt latency.


Soft real-time tasks with less stringent safety requirements run in the \gls{hostd}, based on the Cheshire platform~\cite{ottavianoCheshireLightweightLinuxCapable2023}, extended with a dual-core RV64GCH processor with hardware-assisted virtualization capabilities: it supports concurrent execution of RTOS and GPOS \glspl{vg} through \gls{rv}-compliant H-extension, and virtual interrupts are managed by per-core \gls{vclic} to reduce interrupt handling and context-switching latency among \glspl{vg}. 
The \gls{hostd} cores have private 32KiB L1 \glspl{dc} and share a 128KiB \gls{dpllc}, which interfaces with two external HyperRAM chips via a 400Mb/s deterministic access time HyperBUS memory controller. 
The system interconnect is based on a 64b AXI4 bus. A 1MiB on-chip shared L2 \gls{dcspm} is accessible by all domains with 128b/cyc bandwidth. The SoC also includes conventional peripherals, as shown in Fig.~\ref{fig:soc-archi}. Compute-intensive \gls{ai}/\gls{dsp} workloads are offloaded to two domain-specific accelerators: the vector and \gls{amr} clusters. 

\subsubsection{Hardware IPs For Time-Predictability}

\glspl{mct} on the SoC can interfere through the interconnect and memory endpoints - L2 \gls{dcspm} and HyperRAM accessed via the \gls{dpllc}, resulting in non-deterministic behavior and significantly increasing execution time of \glsplf{tct}, as detailed in \looseness=-1 Sec.~\ref{sec:interference-free}.

To ensure predictable communication over AXI4, each initiator is equipped with a software programmable \glsf{tsu}, shown in Fig.~\ref{fig:hw-ips-for-predictability}.a), aimed at reducing execution latency of \glspl{tct} in interference scenarios by controlling each initiator's bandwidth, thereby enforcing a configurable latency upper bound. The \gls{tsu} comprises three components: 
\textbf{1)} The \textit{\gls{gbs}} fragments long AXI4 bursts to a configurable size to ensure fair arbitration between asynchronous initiators with burst capabilities running \glspl{nct} (e.g., \gls{dma} engines paired with \glspl{dsa}) and initiators running higher-priority \glspl{tct}; 
\textbf{2)} The \textit{\gls{wb}} buffers \emph{AW} and \emph{W} channels, forwarding \emph{AW} requests and \emph{W} bursts only when write data is fully within the buffer. This prevents an initiator from holding the \emph{W} channel, avoiding interconnect stalls; 
\textbf{3)} The \textit{\gls{tru}} assigns each initiator a fixed transfer budget within a configurable communication period. 

Favoring \glspl{tct} in the interconnect unavoidably affects the performance of \glspl{nct}. However, \glspl{mct} conflicting on L2 or HyperRAM endpoints can be further isolated by creating interference-free memory paths. 
\gls{dcspm} can be accessed via two AXI4 ports and addressed in contiguous mode to isolate its physical banks. This is configurable at runtime with zero additional latency through aliased memory map addresses, as shown in Fig.~\ref{fig:hw-ips-for-predictability}.b). For \glspl{nct} sharing L2 data, the \gls{dcspm} operates in interleaved mode to statistically minimize conflicts. 
For \glspl{mct} accessing HyperRAM, the \gls{dpllc} reduces non-deterministic cache misses by creating set-based spatial partitions of configurable sizes, isolated in hardware and assigned to \glspl{vg}' tasks via \textit{part\_id} identifiers linked to AXI4 user signals, as shown in Fig.~\ref{fig:hw-ips-for-predictability}.c). Predictable cache states associated with tasks sharing a partition are maintained by selective partition flushing, preserving the isolation of other \looseness=-1 partitions. 

\subsubsection{Compact, Efficient, \gls{rv} Vector Cluster}
\label{sec:vector}

The proposed SoC integrates a cluster (Fig.~\ref{fig:soc-archi}, top-right) of two compact, energy-efficient \glspl{rvu} controlled by two 32b \gls{rv} scalar cores, forming a clock-gatable \gls{cc}. 
To minimize access energy on the \gls{vrf}, each \gls{rvu} instantiate 2KiB latch-based private \gls{vrf}, connected to the 16-banks, 1024b/cyc-bandwidth L1 \gls{spm}, via four independent 64b \gls{vlsu} ports and a low-latency interconnect. This allows vector engines to quickly perform unit-strided, non-unit-strided, and indexed memory accesses, improving compute efficiency for both dense and sparse workloads. A third \gls{rv} core in the cluster manages a 512b/cyc read/write DMA for double-buffered L2-L1 transfers. 

Each \gls{rvu} is a VLEN=512b RVZve64d processor, supporting formats from FP8 to FP64, bfloat16, integer, and mixed-precision, including \gls{sdotp} operations. Instructions are fetched by the scalar core, decoded by the vector controller to determine the vector length and element width, and executed by the \gls{vau}, achieving a maximum throughput of 256b/cyc for \gls{fp} operations. The \gls{vrf} is organized in four banks, each with 3 read and 1 write 256b ports to meet 3x256b/cyc input, 256b/cyc output bandwidth needs of the \emph{vfmacc} instruction. 
The vector cluster achieves 97.9\% \gls{fpu} utilization at 15.67 DP-FLOP/cyc on edge-sized \glspl{matmul}, up to 121.8 FLOP/cyc on FP8xFP8 \glspl{matmul}, improving performance by 23.8$\times$ to 190.3$\times$ over the \gls{hostd}.

\subsubsection{\gls{amr} Cluster for Mission-Critical \gls{ai}}
\label{sec:int_cluster}
\begin{figure}[t]
    \centering
    \includegraphics[width=0.98\columnwidth]{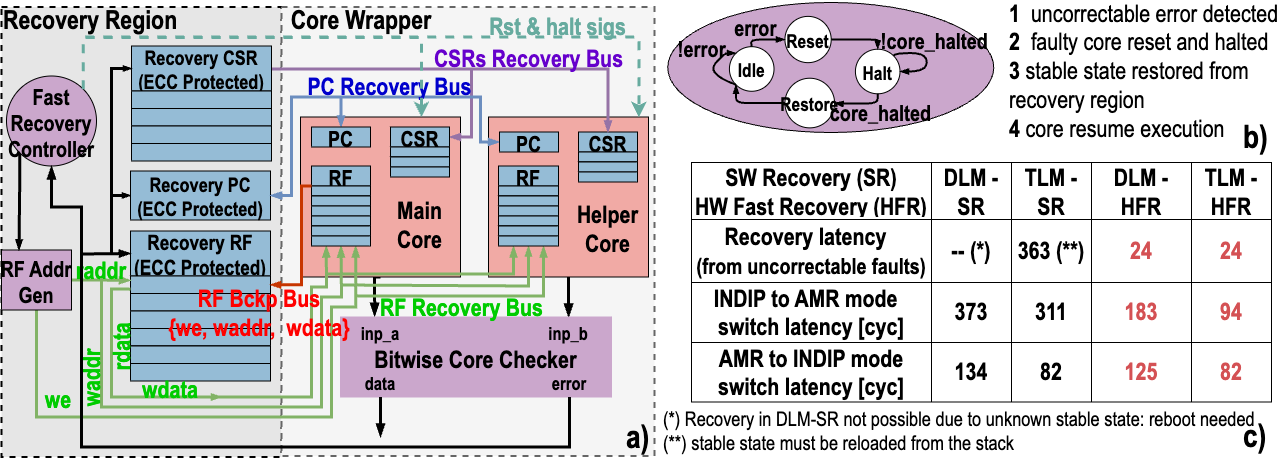}
    \caption{a) Hardware fast recovery (HFR) mechanisms for dual-lockstep mode (DLM) case. b) HFR Finite State Machine. c) \gls{amr} performance.}
    \label{fig:fast-recovery-cluster}
\end{figure}

The \glsf{amr} cluster, shown in bottom right of Fig.~\ref{fig:soc-archi}, includes 12 RV32IMFC cores sharing a 32-banked 256KiB \gls{ecc}-protected L1 \gls{spm}, accessible through a one-cycle latency interconnect with 921Gb/s (@ 900MHz) bandwidth. A 64b/cyc read, 64b/cyc write DMA enables double-buffered L2-L1 data transfers. Mixed-precision integer \gls{dsp}/\gls{ai} tasks are accelerated by the cores through custom \gls{rv} extensions supporting SIMD \gls{sdotp} on data formats ranging from 16b to 2b (all possible mixed permutations). A custom \emph{mac-load} instruction increases MAC unit utilization to 94\% on \glspl{matmul}, overlapping \gls{sdotp} operations with load instructions. 

The RV32 cores can be reconfigured through the \gls{amr} hardware to prioritize reliable vs. more performant execution. In the \gls{indip}, all 12 cores operate in MIMD for maximum performance. In \glsf{dlm} /\gls{tlm} mode, six/four main cores have one/two shadow cores, and instructions are committed after a checker/voting mechanism if no error occurs. The \gls{amr} is runtime-programmable; reconfiguration among modes takes 82-183 clock cycles when switching from safety-critical to high-performance sections within application codes (Fig.~\ref{fig:fast-recovery-cluster}.c)). 

In case of errors, faulty cores are restored to the nearest reliable state in as few as 24 clock cycles thanks to the \glsf{hfr}, shown in Fig.~\ref{fig:fast-recovery-cluster}.a). 
The \gls{hfr} includes \gls{ecc}-protected recovery registers to back up the internal state of non-faulty cores cycle-by-cycle without extra latency. 
As shown in Fig.~\ref{fig:fast-recovery-cluster}.b), \gls{dlm} with \gls{hfr} prevents rebooting the cluster upon fault detection, while \gls{tlm} with \gls{hfr} is 15$\times$ faster than \gls{tlm} software recovery. 
When executing in \gls{dlm} (\gls{tlm}), the performance penalty is limited to 1.89$\times$ (2.85$\times$) compared to \gls{indip}, still achieving 23.1 MAC/cyc (15.3 MAC/cyc) \looseness=-1 on 8b \glspl{matmul}.


\section{Evaluation and Measurements}
\label{sec:evaluation-measurements}
\begin{figure}[t]
    \centering
    \includegraphics[width=0.98\columnwidth]{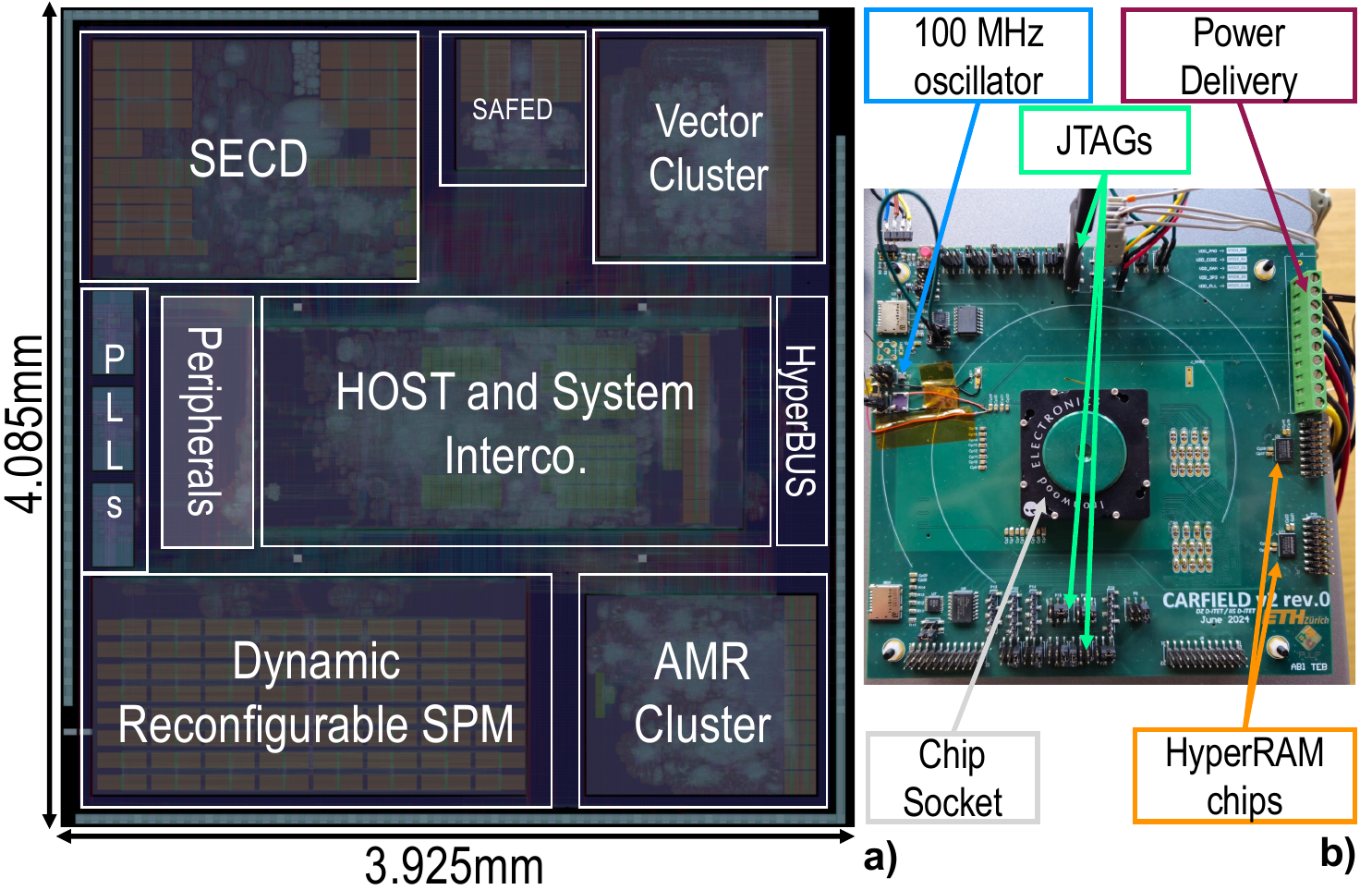}
    \caption{a) SoC micro-graph; b) testing setup. }
    \label{fig:chip-die}
\end{figure}
\begin{figure}[t]
    \centering
    \includegraphics[width=0.98\columnwidth]{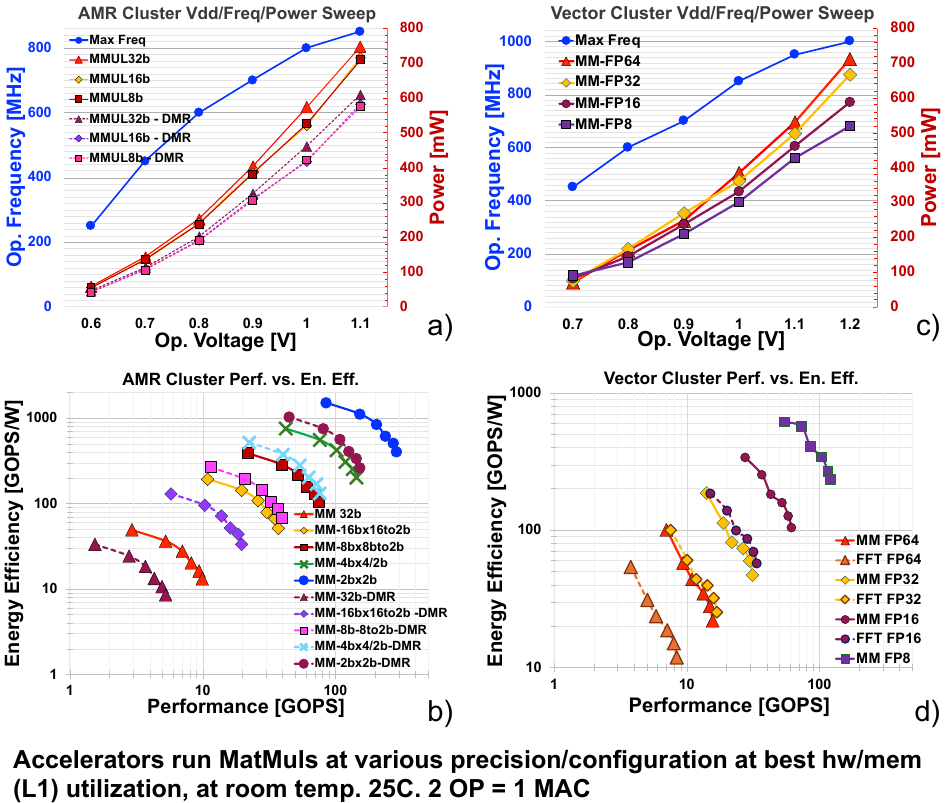}
    \caption{Voltage/Frequency/Power and Performance/Energy Efficiency sweeps of \gls{amr}  (\textbf{a}, \textbf{b}) and vector (\textbf{c}, \textbf{d}) clusters.}
    \label{fig:clus-sweeps2}
\end{figure}
\begin{figure}[t]
    \centering
    \includegraphics[width=.98\columnwidth]{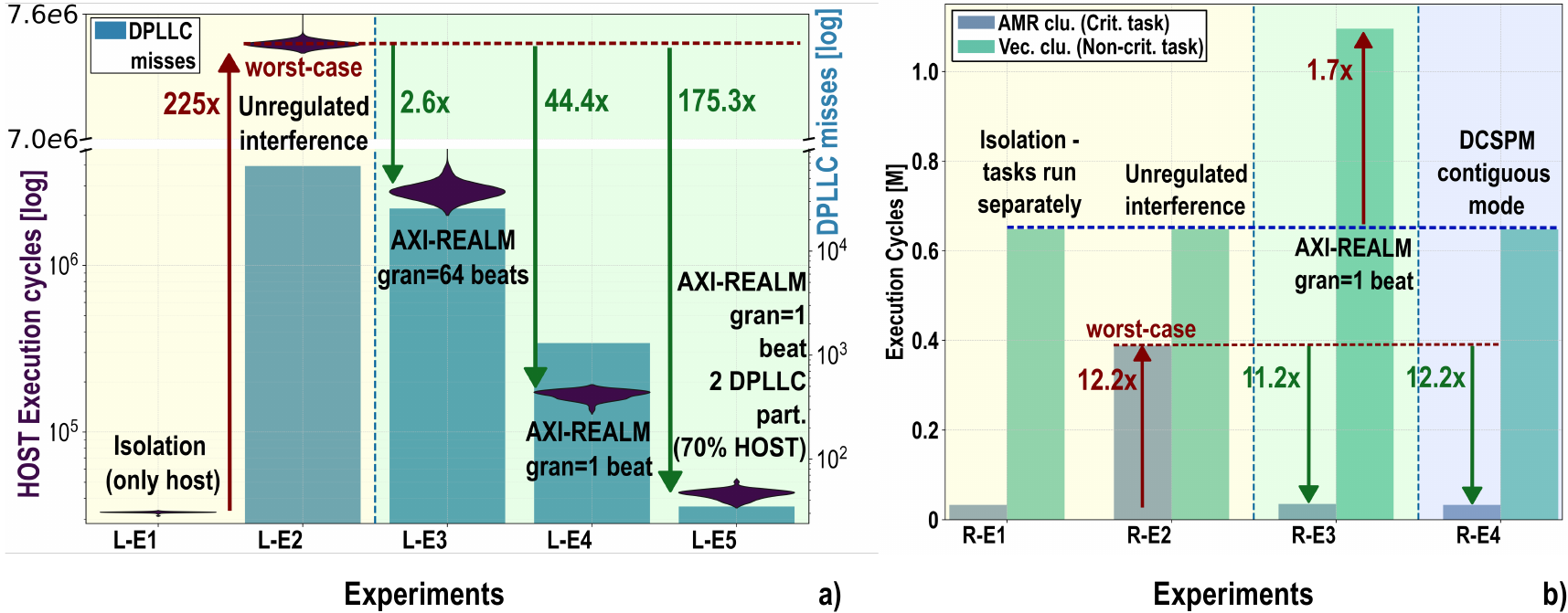}
    \caption{Inteference-aware execution of \glspl{mct} on the SoC: a) \gls{hostd} runs a \gls{tct} accessing HyperRAM, while vector cluster interferes; b) \glspl{mct} running on \gls{amr} and vector clusters in double-buffering, sharing AXI and \gls{dcspm} resources.}
    \label{fig:interference-free}
\end{figure}

Fig.~\ref{fig:chip-die} shows the chip micrograph and the standalone \gls{pcb} designed for testing it. The SoC, fabricated with Intel 16nm FinFet technology, operates from 0.6V to 1.1V, with an overall 1.2W power envelope at a nominal 0.8V. 


\subsubsection{Performance vs Energy Efficiency}
\label{sec:perf-vs-ee}
Fig.~\ref{fig:clus-sweeps2} evaluates the two clusters, in terms of performance and energy efficiency. Measurements are taken at the maximum operating frequency, sweeping the supply voltage from 0.6V to 1.1V. 

The \gls{amr} cluster is benchmarked on integer \glspl{matmul}, the core kernel of \gls{dsp} and \gls{ai} tasks, spanning all supported operands' precisions, from 32b down to 2b, including mixed-precision formats. The cluster's cores are configured to run either in \gls{indip} mode for the best performance or in \gls{dlm} mode for a good trade-off between performance and reliability. The \gls{amr} cluster achieves up to 304.9 GOPS on 2b$\times$2b \glspl{matmul} (161.4 GOPS in \gls{dlm}) at 1.1V, 900 MHz, with a peak energy efficiency of 1.6 TOPS/W at 0.6V, 300 MHz (1.1 TOPS/W in \gls{dlm}).
Similarly, we benchmark the vector cluster on FP \glspl{matmul} and \gls{ffts}, spanning all supported precisions, reaching a peak performance of 122 GFLOPS on FP8 \glspl{matmul} at 1.1V and 1GHz, and a peak energy efficiency of 1.1 TOPS/W at 0.6V and 250 MHz.

\subsubsection{Interference-aware \glspl{mct} execution}
\label{sec:interference-free}
Fig.~\ref{fig:interference-free} shows how the hardware IPs introduced in Sec.~\ref{fig:hw-ips-for-predictability} enable interference-aware execution of \glsplf{mct}. 
In Fig.~\ref{fig:interference-free}.a) the \gls{hostd} runs a \glsf{tct} accessing HyperRAM via the \gls{dpllc} with contiguous stride, while the system DMA interferes by asynchronously transferring data from HyperRAM to the \gls{dcspm} with linear bursts. The measurements show the task latency and jitter, as well as the \gls{dpllc} misses generated by the eviction of cache lines among interfering tasks. In unregulated interference, the \gls{tct} latency degrades by 225$\times$ compared to the isolated (no interference) case (i.e. no interfering DMA). Tuning the \textit{granular burst splitter} and the \textit{traffic regulation unit} of the \glsf{tsu} in software, we regulate the traffic on the interconnect, reducing the latency by 44.4$\times$ compared to the unregulated case. The \gls{tsu} incurs an additional latency of at most 1 clock cycle due to its write buffer. Moreover, by assigning $> 50\%$ spatial partition of the \gls{dpllc} to the \gls{tct}, we reduce cache misses, achieving 75\% of the isolated (no interference) performance.

In a second scenario, the \gls{amr} cluster executes a compute-intensive \gls{tct} in reliable mode, and the vector cluster interferes by executing a \gls{fp} \gls{matmul}. Both accelerators move data in double-buffering from L2 to private L1, overlapping data transfer and computation phases. 
In Fig.~\ref{fig:interference-free}.b), R-E2, the performance of \gls{amr} cluster drops by 12.2$\times$ due to conflicts generated by the vector cluster on the interconnect and the \gls{dcspm}. Programming the \gls{tsu} to regulate the traffic in favor of the \gls{amr} cluster, we reach 95\% of its isolated (no interference) performance (R-E3), degrading the performance of \glsplf{nct}. However, using aliased addresses to access the \gls{dcspm}, we create private memory paths (at zero extra performance overhead) and achieve interference-free execution, matching the isolated (no interference) performance for both tasks \looseness=-1 (R-E4).

We show that these hardware IPs ensure interference-aware concurrent execution of \gls{mct} on shared SoC resources, prioritizing \glspl{tct} over \glspl{nct}  \looseness=-1 with negligible performance overhead.

\section{Comparison with State-of-the-Art}

\begin{figure}[t]
    \centering
    \includegraphics[width=0.98\columnwidth]{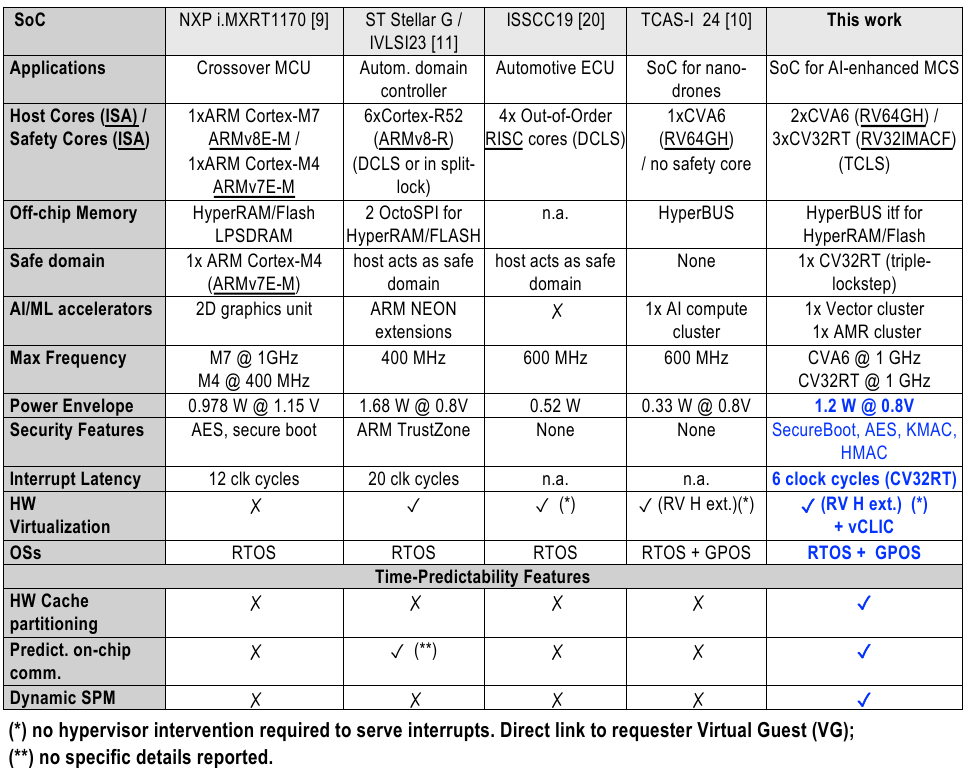}
    \caption{Comparison against \gls{soa} heterogeneous SoC for \gls{mcs}.}
    \label{fig:soa-safety}
\end{figure}

\subsection{Comparison with Mixed-Criticality SoCs}
\label{sec:mcs-soa}

Fig.~\ref{fig:soa-safety} compares the proposed SoC against \gls{cots} and \gls{soa} heterogeneous SoC prototypes presented in the literature in the same power class. 
\gls{cots} solutions like the \textit{I.MXRT1170} crossover MCUs by NXP~\cite{nxp_industrial_control} mainly address heterogeneous workloads for the average case, relying on ARM Cortex-M cores and \gls{gp} acceleration units. However, they lack hardware mechanisms for reliability, virtualization, and time predictability, limiting their use as a \glspl{mcs}. Renesas' prototype~\cite{otani2728nm600MHz2019} focuses on reliable and virtualized RISC cores, reducing \glspl{vg} context-switch overheads, but it does not enable resource partitioning among concurrent \glspl{vg} in hardware, nor does it provide \gls{ai}/\gls{dsp} hardware acceleration. Another example is the \textit{Stellar} processor by ST~\cite{grossierASILDAutomotivegradeMicrocontroller2023}. Despite integrating reliable and real-time ARM Cortex-R52 cores operating in split-lock and an interconnect with Quality of Service mechanisms, it lacks hardware IPs to observe and partition shared memory endpoints. Moreover, it relies solely on \gls{gp} cores for compute-intensive tasks. 
In academia,~\cite{valenteHeterogeneousRISCVBased2024} targets \gls{ai}-enhanced nano-drone applications with a SoC featuring a single-core 64b \gls{rv} processor and an acceleration cluster similar to ours. However, it lacks features for reliability and time-predictability, limiting the~SoC usability in mission-critical scenarios. 

At a comparable power envelope, the proposed SoC is the only one that integrates comprehensive hardware mechanisms for time predictability, enabling software-controlled (e.g., via hypervisors) dynamic partitioning of shared resources like interconnects, caches, and \gls{spm}. It achieves the fastest interrupt latency response, 2$\times$, 3.3$\times$, and 8.3$\times$ lower than~\cite{nxp_industrial_control}, \cite{grossierASILDAutomotivegradeMicrocontroller2023}, and~\cite{valenteHeterogeneousRISCVBased2024}, respectively. Additionally, it supports concurrent execution of GPOS and RTOS, integrates a HW RoT secure domain with extensive security primitives, and a comprehensive set of programmable accelerators for compute-intensive mixed-criticality workloads, achieving \gls{soa} performance and energy \looseness=-1  efficiency.

\subsection{Comparison with edge \gls{ai} and Vector Processors}
\label{sec:soa-acc}
\begin{figure}[t]
    \centering
    \includegraphics[width=0.98\columnwidth]{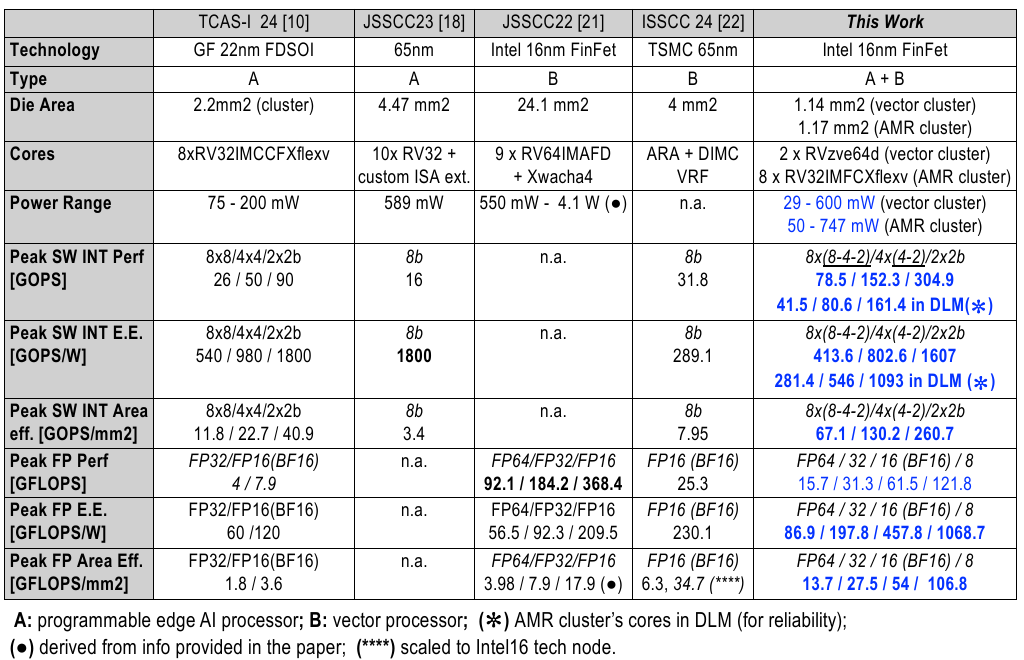}
    \caption{Comparison against \gls{soa} accelerators for \gls{fp} and edge \gls{ai}.}
    \label{fig:soa-accelerators}
\end{figure}

Fig.~\ref{fig:soa-accelerators} compares the proposed \gls{amr} and vector clusters against \gls{soa} vector and edge \gls{ai} processors. The \gls{amr} cluster compares favorably to a similar parallel cluster that supports low-bit-width integer arithmetic but lacks reliability features~\cite{valenteHeterogeneousRISCVBased2024}, limiting its use in mission-critical applications. In \gls{dlm}, it achieves up to 1.8$\times$ better performance (3.4$\times$ in \gls{indip}) on uniform 8b/4b/2b \glspl{matmul}, with  6.4$\times$ better area efficiency and comparable energy efficiency. In \gls{dlm}, it provides 2.6$\times$ higher performance than the most efficient 8b integer processor~\cite{juSystolicNeuralCPU2023}, which however lacks support for reliable execution modes. 
On \gls{fp} workloads, our vector cluster is the only one that operates over the full range of \gls{fp} formats, from 64b down to 8b, achieving the highest computing efficiency due to near-ideal resource utilization. Compared to~\cite{schmidtEightCore144GHzRISCV2022}, implemented in the same technology node, our cluster demonstrates 2.2$\times$ and 3$\times$ higher energy and area efficiency, respectively, on FP16 workloads. 
Additionally, it shows 2.43$\times$ better performance, 2$\times$, and 1.6$\times$ higher energy and area efficiency than~\cite{wang306Vecim28913GOPS2024}, despite the latter leveraging compute-in-memory in the VRF.


\section{Conclusion}
 In this brief, we presented a 16nm reliable, time-predictable heterogeneous RISC-V SoC for \gls{ai}-enhanced mixed-criticality applications. To the best of our knowledge, this is the first SoC that combines safety features for \glspl{mcs} with hardware IPs for time-predictable execution of \glspl{mct} and leading-edge domain-specialized programmable accelerators within the same heterogeneous SoC. With a peak performance of 304.9 GOPS at 1.6TOPS/W and 260.7 GOPS/mm$^2$, and 121.8 GFLOPS at 1.1TFLOPS/W and 107GFLOPS/mm$^2$, the proposed SoC offers a comprehensive solution for reliable and deterministic execution of \gls{ai}/\gls{dsp}-enhanced \gls{mc} edge applications, achieving \gls{soa} energy efficiency under \looseness=-1  1.2~W power envelope. 
%


\section*{Acknowledgments}
This work was supported by the HORIZON CHIPS-JU TRISTAN (101095947) and ISOLDE (101112274) projects.

\bibliographystyle{ieeetr}
\bibliography{REFERENCES/carfield_soc_bib_brev}

\end{document}